\documentclass[letterpaper,iop]{emulateapj}
\pdfoutput=1
\usepackage{thumbpdf} 
\usepackage{amsmath,amssymb} 
\usepackage{graphicx,mathptmx} 
\usepackage{natbib} 
\usepackage[usenames]{color} 
\usepackage{ifpdf}  
\usepackage{xcolor}[2007/01/21] 
\usepackage{refcount} 
\usepackage{textcase} 
\usepackage{epstopdf} 

\ifpdf
\pdfoutput=1 
\else
\fi

\newcommand{\beq}{
\begin{equation}
}
\newcommand{\eeq}{
\end{equation}
}
\newcommand{\beqa}{
\begin{eqnarray}
}
\newcommand{\eeqa}{
\end{eqnarray}
}

\newcommand{\msun}     {\ensuremath{{{M}}_{\scriptscriptstyle \odot}}}

\newcommand{\kms}      {\ensuremath{~\mathrm{km~s^{-1}}}}

\newcommand{\Mpc}      {\ensuremath{~\mathrm{Mpc}}}
\newcommand{\msigma}   {\ensuremath{M}{--}\ensuremath{\sigma}}

\newcommand{\mbh}      {\ensuremath{M}}

\providecommand{\ion}[2]{#1$\;$\textsmaller{\@Roman{#2}}}
\newcommand{\mumax}      {\ensuremath{\mu_{\mathrm{max}}}}
\newcommand{\mumin}      {\ensuremath{\mu_{\mathrm{min}}}}

\newcommand{\half}{{\textstyle{\frac{1}{2}}}}

\def\spose#1{\hbox to 0pt{#1\hss}}
\newcommand{\lta}{\mathrel{\spose{\lower 3pt\hbox{$\mathchar"218$}}
      \raise 2.0pt\hbox{$\mathchar"13C$}}}
\newcommand{\gta}{\mathrel{\spose{\lower 3pt\hbox{$\mathchar"218$}}
      \raise 2.0pt\hbox{$\mathchar"13E$}}}
\def\simlt{\mathrel{\rlap{\lower 3pt\hbox{$\sim$}}\raise 2.0pt\hbox{$<$}}}
\def\simgt{\mathrel{\rlap{\lower 3pt\hbox{$\sim$}} \raise 2.0pt\hbox{$>$}}}

\definecolor{KayhanCiteColor}{rgb}{0,0.08,0.35}
\definecolor{KayhanURLColor}{rgb}{0,0.08,0.35}
\definecolor{KayhanLinkColor}{rgb}{0,0.08,0.35}
\definecolor{KayhanPageColor}{rgb}{0,0.08,0.35}
\definecolor{medred}{rgb}{0.75,0.0,0.0}

\shorttitle{Selection effects on $M$--$\sigma$ relation}
\shortauthors{G\"{u}ltekin et al.}

\submitted{Accepted by The Astrophysical Journal}

\makeatletter
\ifx\undefined\hyperref
\usepackage[pdftitle={Observational selection effects and the M-sigma relation},pdfauthor={Kayhan Gultekin},pdfsubject={Astrophysics},pdfkeywords={black hole physics --- galaxies: general --- galaxies: nuclei --- galaxies: bulges --- methods: statistical},letterpaper,linktocpage,colorlinks=true,linkcolor={KayhanLinkColor},urlcolor={KayhanURLColor},citecolor={KayhanCiteColor},hyperfootnotes=false]{hyperref}
\else\relax\fi
\makeatother
\usepackage[figure,figure*]{hypcap} 
\usepackage{atveryend} 
\usepackage[
  open,
  openlevel=3,
  atend
]{bookmark}[2010/04/08]  
\begin{document}

\label{firstpage}
 
\title{Observational selection effects and the\ $M$--$\sigma$ relation}

\author{Kayhan G\"{u}ltekin}
\affil{Department of Astronomy, University of Michigan, Ann Arbor, MI 48109
\href{mailto:kayhan@umich.edu}{kayhan@umich.edu}.}
\author{Scott Tremaine}
\affil{School of Natural Sciences, Institute for Advanced Study,
 Princeton, NJ 08540}
\author{Abraham Loeb}
\affil{Astronomy Department, Harvard University, Cambridge, MA 02138}
\author{Douglas O.\ Richstone}
\affil{Department of Astronomy, University of Michigan, Ann Arbor, MI 48109}

\begin{abstract}
\hypertarget{abstract}{}%
We examine the possibility that the observed relation between
black-hole mass and host-galaxy stellar velocity dispersion (the
\msigma\ relation) is biased by an observational selection effect, the
difficulty of detecting a black hole whose sphere of influence is
smaller than the telescope resolution.  In particular, we critically
investigate recent claims that the \msigma\ relation only represents
the upper limit to a broad distribution of black-hole masses in
galaxies of a given velocity dispersion. We find that this hypothesis
can be rejected at a high confidence level, at least for the early-type galaxies
with relatively high velocity dispersions (median $268\ \kms$) that
comprise most of our sample. We also describe a general
procedure for incorporating observational selection effects in
estimates of the properties of the \msigma\ relation. Applying this
procedure we find results that are consistent with earlier estimates
that did not account for selection effects, although with larger error
bars. In particular, (\emph{i}) the width of the \msigma\ relation is
not significantly increased; (\emph{ii}) the slope and normalization
of the \msigma\ relation are not significantly changed; (\emph{iii})
most or all luminous early-type galaxies contain central black holes
at zero redshift.  Our results may not apply to late-type or small
galaxies, which are not well-represented in our sample. 

\bookmark[ rellevel=1, keeplevel,
dest=abstract
]{Abstract}
\end{abstract}

\section{Introduction}
\label{intro}

The mass of a central black hole is correlated with the properties of
its host galaxy, both stellar luminosity or mass \citep{dressler89,
  kormendy93a, magorrianetal98, 2004ApJ...604L..89H} and velocity
dispersion \citep[the \msigma\ relation;][hereafter
G09]{gebhardtetal00a, fm00,2009ApJ...698..198G}. These
correlations provide deep, but poorly understood, insights into galaxy
and black-hole formation (see \S\ref{sec:disc} for a brief
review). The radius of the sphere of influence of a central black hole
of mass \mbh\ in a galaxy with velocity dispersion $\sigma$ is
$R_\mathrm{infl} \equiv G \mbh \sigma^{-2}$, so at a distance $D$ the
angular size of the sphere of influence is $\theta_\mathrm{infl} =
R_\mathrm{infl}/D$.  An important determinant of the reliability of
dynamical detections of central black holes is the ratio of
the radius of the sphere of influence to the telescope resolution.
Thus we desire $\theta_\mathrm{infl}\gtrsim \theta_\mathrm{res}$ where
$\theta_\mathrm{res}$ is some measure of the angular resolution, for
example the slit width or the full width at half maximum (FWHM) of the
telescope point-spread function.  With few exceptions, the black holes
in nearby galaxies have $\theta_\mathrm{infl}\lesssim 1\arcsec$ (see
Table 1)---this is why most detections so far have been made with the
\emph{Hubble Space Telescope} (\emph{HST}), which offers higher
spatial resolution (FWHM$\simeq 0\farcs1$) and a more stable
point-spread function than ground-based telescopes.

At a given signal-to-noise ratio, as the ratio $\theta_\mathrm{infl} /
\theta_\mathrm{res}$ decreases the measurement errors in black-hole
mass increase until the black-hole mass becomes consistent with zero.  Since
$\theta_\mathrm{infl}$ is usually not much bigger than
$\theta_\mathrm{res}$, an obvious concern is that resolution-dependent
selection effects may bias the observed correlations between
black-hole mass and galaxy properties.  We focus here on possible bias
in the \msigma\ relation, although similar considerations apply to the
relations between black-hole mass and host galaxy luminosity or mass.
A number of possible biases have been discussed in the literature.
Several authors have argued that black-hole masses are systematically
overestimated when the sphere of influence is not well-resolved (see
\S4.1 of G09 and references therein).  This seems unlikely for the
following reasons: (\emph{i}) Assuming that the experimental analyses of
black-hole mass measurements are properly designed, the model parameters
derived from poor data may have large error bars but should not be
systematically biased.  (\emph{ii}) \cite{gebhardtetal03} analyzed 12
galaxies twice, once using both ground-based (low-resolution) and
\emph{HST} (high-resolution) spectroscopy and once using only the
ground-based spectroscopy.  They found that the black-hole masses
determined from these two data sets were consistent at the 1-$\sigma$
level, with no evidence that the masses determined from ground-based
data alone were systematically high.  (\emph{iii}) \cite{kormendy04}
has pointed out that the black-hole mass in M32 (NGC 0221) has
remained remarkably stable---within a factor of two---over the past
two decades while the spatial resolution of the spectroscopy has
increased by a factor of 30. Kormendy also argues that the mass estimates for several
black holes first marginally resolved from the ground (e.g., NGC 3115, NGC
3377, NGC 4594) did not systematically change when they were later
observed with much higher resolution by \emph{HST}---the error bars
shrank but the best-estimate mass did not change significantly.  In fact, G09
pointed out a different and more important bias that is the opposite
of this one: \emph{excluding} black-hole masses from galaxies with
$\theta_\mathrm{infl}/\theta_\mathrm{res}\lesssim 1$ from a mass
sample systematically biases the \msigma\ relation derived from that
sample.

Yet another bias occurs when non-detections of black holes (i.e.,
measured upper limits to the black-hole mass) are not included in the
analysis: if the upper limits are not far from the ridgeline of the
\msigma\ relation, then analyses that exclude them will be biased
towards high mass at a given dispersion.  On the other hand if upper
limits are included, the analysis must account for the possibility
that some galaxies do not contain black holes at all, or else a single
galaxy without a black hole could drastically alter the best-fit
parameters of the \msigma\ relation.  Most studies have considered
only measured black-hole masses and have ignored the numerous upper
limits available in the literature; for exceptions see
\cite{vallurietal05} and G09.

One possibility \citep{1999ASSL..234..157H, batch} is that the
\msigma\ relation only describes an upper limit to the black-hole mass
in a host galaxy with given dispersion; that is, the black hole may
have any mass at or below the mass given by the \msigma\ relation.  In
this view, the apparent narrow width of the relation is an
observational selection effect that arises because black holes with
much smaller masses, though common, generally cannot be detected.

This paper has two related goals: (\emph{i}) to determine
quantitatively whether the \msigma\ relation is a {\em ridgeline}
(i.e., most black-hole masses lie close to the relation) or an {\em
upper envelope} (i.e., most black-hole masses lie well below the relation);
(\emph{ii}) to investigate whether the parameters of the \msigma\
relation are biased by the inability to detect black holes when the
angular size of the sphere of influence is too small.  A by-product of
the analysis will be an estimate of the fraction of luminous galaxies
that contain black holes.  In \S\ref{restate} we review the arguments
by \citet{batch} in favor of the upper-envelope model.  In \S\ref{tests} we
test the predictions of the upper-envelope model and test for bias due to
selection effects in the \msigma\ relation.  We find that we can
reject upper-envelope models at very high significance.  \S\ref{sec:disc}
contains a discussion and conclusions.

\section{Review of the upper-envelope model}

\label{restate}

We have argued that detecting black holes is difficult if
$\theta_\mathrm{infl}\lesssim \theta_\mathrm{res}$.  This criterion is
oversimplified, since the ability to measure black-hole mass depends
on a number of factors in addition to angular resolution; some of
these factors depend on the quality of the observations (e.g., the
signal-to-noise ratio) while others depend on the properties of the
galaxy (e.g., detecting black holes in galaxies with large cores is
more difficult than in power-law galaxies)---see G09 for a more
detailed discussion.  However, it is instructive to make the
simplifying assumption that a black hole can be detected only
if $\theta_\mathrm{infl}$ exceeds a fixed fraction of
$\theta_\mathrm{res}$.  In this case, if galaxies are uniformly
distributed in space and the number density of galaxies with given
black-hole mass $\mbh$ and dispersion $\sigma$ is $n(\mbh,\sigma)$,
the number of detected black holes with these parameters will be
proportional to $\mbh^3 n(\mbh,\sigma)$. The additional factor of
$\mbh^3$ can create a bias such that galaxies of a given dispersion
with low-mass black holes are strongly under-represented in the
samples used to determine the \msigma\ relation.

\citet{batch} quantifies this argument using a sample of $\sim 2500$
galaxies with distance $D<100\Mpc$ and measured velocity dispersion,
taken from the HyperLeda catalog\footnote{See
  \href{http://leda.univ-lyon1.fr}{http://leda.univ-lyon1.fr}.}
\citep{hyperleda}. He assigns a black hole to each galaxy, with mass
chosen uniformly random in $\log\mbh$ between a lower limit
$\mbh=10\,\msun$ and an upper limit given by the \msigma\ relation,
and he assumes that the black holes can be detected only
if\footnote{This analysis is also repeated with
  $\theta_\mathrm{res}=0\farcs05$, and yields consistent results.}
$\theta_\mathrm{infl}>\theta_\mathrm{res}=0\farcs1$. The resulting
simulated dataset of black-hole masses yields an apparent \msigma\
relation with scatter comparable to the observed relation; Batcheldor
thus argues that the upper-envelope model is consistent with the data.

This argument implies that \emph{HST} observations should yield many
more upper limits to black-hole mass than actual detections.
Consider, for example, the dispersion range $325\kms < \sigma <
385\kms$, which in the sample of G09 contains four galaxies \citep[IC
1459, NGC 1399, NGC 4486, and NGC 4649, measured
by][respectively]{2002ApJ...578..787C, 2007ApJ...671.1321G, gt09,
2003ApJ...583...92G}. All of these have black-hole masses
$>5\times10^8\msun$. If $\log M$ is uniformly distributed between 10
and $\sim 3\times 10^9\msun$ in this dispersion range, as suggested by
Batcheldor, then for every black hole with $M>5\times10^8\msun$ there
should be roughly 10 (i.e., $[\log(5 \times 10^8) - \log(10)] /
[\log(3 \times 10^9) - \log(5 \times 10^8)]$) with smaller
masses. Thus the four galaxies in the sample of G09 should be
accompanied by $\sim 40$ galaxies with similar dispersions and
distances but smaller black holes. Since the observations are planned
before knowing the mass of the black hole, there should be 40 galaxies
with smaller black holes that have also been observed, for which the
observations would probably yield only upper limits.  Instead of 40,
there are only 3 upper limits in that range in the literature
\citep[NGC 315, NGC 6861, and NGC 1841;][]{beifiorietal08}.  Thus
published measures of black-hole masses argue against the
upper-envelope model, but the published data may not tell the whole
story.  Some upper limits derived from \emph{HST} observations may not
be in the literature, and the observers may have had other clues
leading to an enhanced success rate (e.g., a rising dispersion curve
from ground-based observations, weak AGN activity, etc.).
Nevertheless, there is little or no positive evidence that supports
the notion that very small black holes are present in galaxies with
such high velocity dispersion.

\section{Tests of the upper-envelope model}
\label{tests}

As we have discussed, a critical test of the upper-envelope model is
whether it correctly predicts the success rate of detecting central
black holes. The challenge in applying this test is that we cannot
model the behavior of observers and time allocation committees, who
determine which galaxies are to be observed. However, the most
promising sites to prospect for black holes are the centers of those
galaxies with the largest values of
$\theta^\mathrm{pred}_\mathrm{infl}\equiv
G\mbh^\mathrm{pred}/(\sigma^2D)$, the angular size of the sphere of
influence determined using the black-hole mass $\mbh^\mathrm{pred}$
predicted by the \msigma\ relation. Thus an objectively defined sample
that provides the sharpest tests of the upper-envelope model is the
set of galaxies with the largest values of
$\theta^\mathrm{pred}_\mathrm{infl}$---the difference between the
ridgeline model and the upper-envelope model is maximized for this
sample.

We have queried the HyperLeda catalog for all galaxies with measured
distance and central velocity dispersion. HyperLeda is not complete in
any sense, but this method mimics the approach used by observers to
identify target galaxies for black-hole searches. For each galaxy we
predicted the black-hole mass using the \msigma\ relation in the form
\begin{equation}
\mbh^\mathrm{pred}(\sigma) = 10^\alpha (\sigma / 200 \kms)^\beta\msun
\label{eq:msig}
\end{equation}
with $\alpha=8.12$ and $\beta=4.24$ from G09.  Using other values of
$\beta$ changes the sample, but our final results are very similar
when using any $\beta$ in the range 3--5 to create the initial
sample. We next computed $\theta_\mathrm{infl}^\mathrm{pred}$ and
sorted the galaxies by this parameter.  We then found the best
available distances and dispersions\footnote{Some of these dispersions
seem implausible to us, e.g., $500\kms$ for NGC 4055, and we are
engaged in a program to remeasure high-dispersion galaxies including
some from this list.}  for the galaxies near the top of the list,
recomputed $\theta_\mathrm{infl}^\mathrm{pred}$ and resorted.  The
galaxies with the top 50 resulting values of
$\theta_\mathrm{infl}^\mathrm{pred}$ are listed in Table
\ref{t:gallist}.  

\begin{deluxetable}{llrrrrrr}
  \footnotesize
  \tablecaption{Top 50 galaxies by predicted angular sphere of influence}
  \tablehead{
     \colhead{Galaxy} &
     \colhead{Type} &
     \colhead{$\sigma$} &
     \colhead{$D$} &
     \colhead{$M^\mathrm{pred}$} &
     \colhead{$\theta_\mathrm{infl}^\mathrm{pred}$} &
     \colhead{$M$} &
     \colhead{Ref.} \\[1ex]
     &
     &
     &
     \colhead{(Mpc)} &
     \colhead{($10^8 \msun$)} &
     \colhead{(\arcsec)} &
     \colhead{($10^8 \msun$)} &
     \colhead{}
  }
  \startdata
N0224 & Sb & 160 &  0.8 &   0.52 & 2.27 & 1.5\phantom{0}& 1 \\
N4649 & E2 & 385 & 16.5 &  22\phantom{.00} & 0.79 & 21\phantom{.00}& 2 \\
N6861 & S0 & 414 & 28.1 &  29\phantom{.00} & 0.54 & $<$15\phantom{.00}& 3 \\
N4486 & E1 & 324 & 17.0 &  10\phantom{.00} & 0.52 & 62.7\phantom{0}& 4 \\
N3998 & S0 & 305 & 14.9 &   8.1\phantom{0} & 0.52 & 2.4\phantom{0}& 5 \\
N1399 & E1 & 337 & 21.1 &  12\phantom{.00} & 0.46 & 5.1\phantom{0}& 6 \\
N4751 & S0 & 349 & 23.5 &  14\phantom{.00} & 0.44 & \dots & \dots \\
N4594 & Sa & 240 & 10.3 &   2.9\phantom{0} & 0.44 & 5.7\phantom{0}& 7 \\
N4472 & E2 & 294 & 17.0 &   6.9\phantom{0} & 0.42 & \dots & \dots \\
N4374 & E1 & 296 & 17.0 &   7.1\phantom{0} & 0.42 & 15\phantom{.00}& 8 \\
N3115 & S0 & 230 & 10.2 &   2.4\phantom{0} & 0.40 & 9.6\phantom{0}& 9 \\
N0221 & E2 &  75 &  0.9 &   0.02 & 0.39 & 0.03& 10 \\
N1332 & S0 & 321 & 22.9 &  10\phantom{.00} & 0.38 & 14.5\phantom{0}& 11 \\
N4143 & SB0 & 271 & 16.0 &   4.9\phantom{0} & 0.37 & $<$1.4\phantom{0}& 12 \\
N5128 & S0/E & 150 &  4.4 &   0.4\phantom{0} & 0.36 & 3.0\phantom{0}& 13 \\
N1161 & S0 & 336 & 27.5 &  12\phantom{.00} & 0.35 & \dots & \dots \\
N3031 & Sb & 143 &  4.1 &   0.33 & 0.34 & 0.8\phantom{0}& 14 \\
N4945 & Sc & 134 &  3.7 &   0.25 & 0.33 & \dots & \dots \\
N4552 & E1 & 254 & 15.4 &   3.7\phantom{0} & 0.33 & $<$19\phantom{.00}& 3 \\
N4526 & SAB0 & 264 & 16.9 &   4.4\phantom{0} & 0.33 & $<$3.2\phantom{0}& 3 \\
N2787 & SB0 & 189 &  7.9 &   1.1\phantom{0} & 0.33 & 0.43& 15 \\
N2293 & SAB0$^\mathrm{pec}$ & 261 & 17.1 &   4.2\phantom{0} & 0.32 & \dots & \dots \\
IC1459 & E4 & 340 & 30.9 &  13\phantom{.00} & 0.32 & 28\phantom{.00}& 16 \\
E137$-$044 & SAB0 & 489 & 69.3 &  60\phantom{.00} & 0.32 & \dots & \dots \\
N3034 & Irr & 130 &  4.0 &   0.22 & 0.29 & \dots & \dots \\
E138$-$005 & SB0$^\mathrm{pec}$ & 349 & 36.1 &  14\phantom{.00} & 0.29 & \dots & \dots \\
N4055 & E? & 500 & 87.5 &  66\phantom{.00} & 0.27 & \dots & \dots \\
N3379 & E0 & 206 & 11.7 &   1.5\phantom{0} & 0.27 & 1.2\phantom{0}& 17 \\
N4365 & E3 & 256 & 20.4 &   3.8\phantom{0} & 0.26 & \dots & \dots \\
N4278 & E1 & 237 & 16.7 &   2.8\phantom{0} & 0.26 & $<$1.8\phantom{0}& 3 \\
N1023 & SB0 & 205 & 12.1 &   1.5\phantom{0} & 0.26 & 0.5\phantom{0}& 18 \\
N5087 & S0? & 283 & 26.2 &   5.9\phantom{0} & 0.25 & \dots & \dots \\
N4406 & E3 & 235 & 17.0 &   2.7\phantom{0} & 0.25 & \dots & \dots \\
N4261 & E2 & 315 & 33.4 &   9.3\phantom{0} & 0.25 & 5.5\phantom{0}& 19 \\
N2663 & E & 291 & 27.5 &   6.6\phantom{0} & 0.25 & \dots & \dots \\
N4621 & E5 & 225 & 17.0 &   2.2\phantom{0} & 0.23 & \dots & \dots \\
N3923 & E4--5 & 257 & 22.9 &   3.9\phantom{0} & 0.23 & \dots & \dots \\
N5062 & S0$^\mathrm{pec}$ & 389 & 60.0 &  23\phantom{.00} & 0.22 & \dots & \dots \\
N4342 & S0 & 225 & 18.0 &   2.2\phantom{0} & 0.22 & 3.6\phantom{0}& 20 \\
N4105 & E3 & 262 & 26.6 &   4.2\phantom{0} & 0.21 & \dots & \dots \\
N1407 & E0 & 272 & 28.8 &   4.9\phantom{0} & 0.21 & \dots & \dots \\
N1270 & E? & 427 & 76.7 &  34\phantom{.00} & 0.21 & \dots & \dots \\
N0253 & SABc & 103 &  3.2 &   0.08 & 0.21 & \dots & \dots \\
N5838 & S0 & 266 & 28.5 &   4.5\phantom{0} & 0.20 & \dots & \dots \\
N6587 & SAB0? & 333 & 48.9 &  12\phantom{.00} & 0.19 & \dots & \dots \\
N1395 & E2 & 245 & 24.6 &   3.2\phantom{0} & 0.19 & $<$0.14& 3 \\
IC2586 & E4 & 346 & 53.1 &  14\phantom{.00} & 0.19 & \dots & \dots \\
N4291 & E2 & 242 & 25.0 &   3\phantom{.00} & 0.18 & 3.2\phantom{0}& 2 \\
N2841 & Sb & 206 & 17.8 &   1.5\phantom{0} & 0.18 & \dots & \dots \\
N4442 & SB0 & 187 & 15.3 &   1\phantom{.00} & 0.17 & \dots & \dots 

\enddata
\label{t:gallist}
\tablecomments{The 50 galaxies with the largest angular sphere of
  influence, as predicted by the \msigma\ relation.  Hubble types are
  mostly from NED.  The velocity dispersions come from G09 or
  HyperLeda, and the distances are our best estimates from G09,
  \citet{tonryetal01}, NED redshift-independent distances, or
  HyperLeda.  Predicted black-hole masses are from equation
  (\ref{eq:msig}), and $\theta^\mathrm{pred}_\mathrm{infl} = G
  M^\mathrm{pred}\sigma^{-2} D^{-1}$.  We also list the best
  black-hole mass measurements for these galaxies.  The final column
  gives a reference code for the mass measurement or upper limit.}
\tablerefs{
(1) \cite{benderetal05}, 
(2) \cite{2003ApJ...583...92G}, 
(3) \cite{beifiorietal08},
(4) \cite{gt09} and \cite{2011arXiv1101.1954G},
(5) \cite{2006AandA...460..439D}, 
(6) \cite{2007ApJ...671.1321G}, 
(7) \cite{1988ApJ...335...40K}, 
(8) \cite{1998ApJ...492L.111B}, 
(9) \cite{1999MNRAS.303..495E},
(10) \cite{2002MNRAS.335..517V}, 
(11) \cite{ruslietal10},
(12) \cite{2002ApJ...567..237S},
(13) \cite{2005AJ....130..406S}, 
(14) \cite{2003AJ....125.1226D}, 
(15) \cite{2001ApJ...550...65S},
(16) \cite{2002ApJ...578..787C}, 
(17) \cite{2000AJ....119.1157G}, 
(18) \cite{2001ApJ...550...75B}, 
(19) \cite{1996ApJ...470..444F}, 
(20) \cite{1999ApJ...514..704C}.
}
\end{deluxetable}
\bookmarksetup{color=[rgb]{0,0,0.54}}
\bookmark[
rellevel=1,
keeplevel,
dest=table.\getrefnumber{t:gallist}
]{Table \ref*{t:gallist}: Top 50 galaxies by predicted angular sphere of influence}
\bookmarksetup{color=[rgb]{0,0,0}}

Our results will be based on a sub-sample of galaxies from this Table
with the $N\le 50$ largest predicted angular spheres of influence, and
we must choose $N$. If $N$ is too small, the statistical uncertainties
will be unnecessarily large. If $N$ is too large the power of the test
will be diluted by galaxies that have not been examined for black
holes. We normally work with $N=30$, but we have experimented with
other values of $N$ and find, as described below, that our results are
quite insensitive to $N$ so long as $N\gtrsim 20$. Of the top 30
galaxies in Table \ref{t:gallist}, 15 have published black-hole mass
determinations and 5 have published upper limits.

We present two tests with these data.  The first tests for the
probability of obtaining these data given the upper-envelope hypothesis as presented
by \citet{batch}.  The second examines more generally how 
limited resolution affects our inferences about the \msigma\ relation and
its properties.

\subsection{Test A}

\begin{figure*}[htpb]
\hypertarget{pvsmu}{}%
\hypertarget{pvssoi}{}%
\hypertarget{pvsdelta}{}%
\centering
\includegraphics[width=0.33\textwidth]{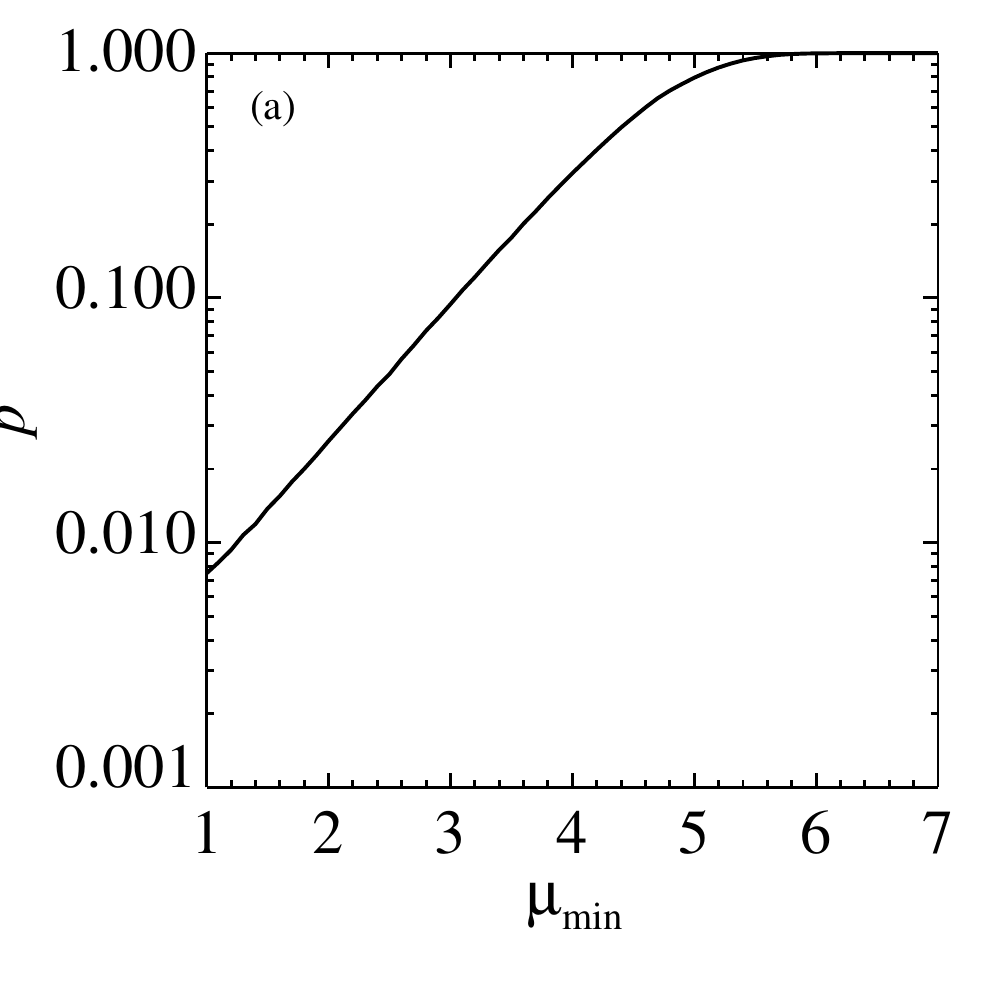}
\includegraphics[width=0.33\textwidth]{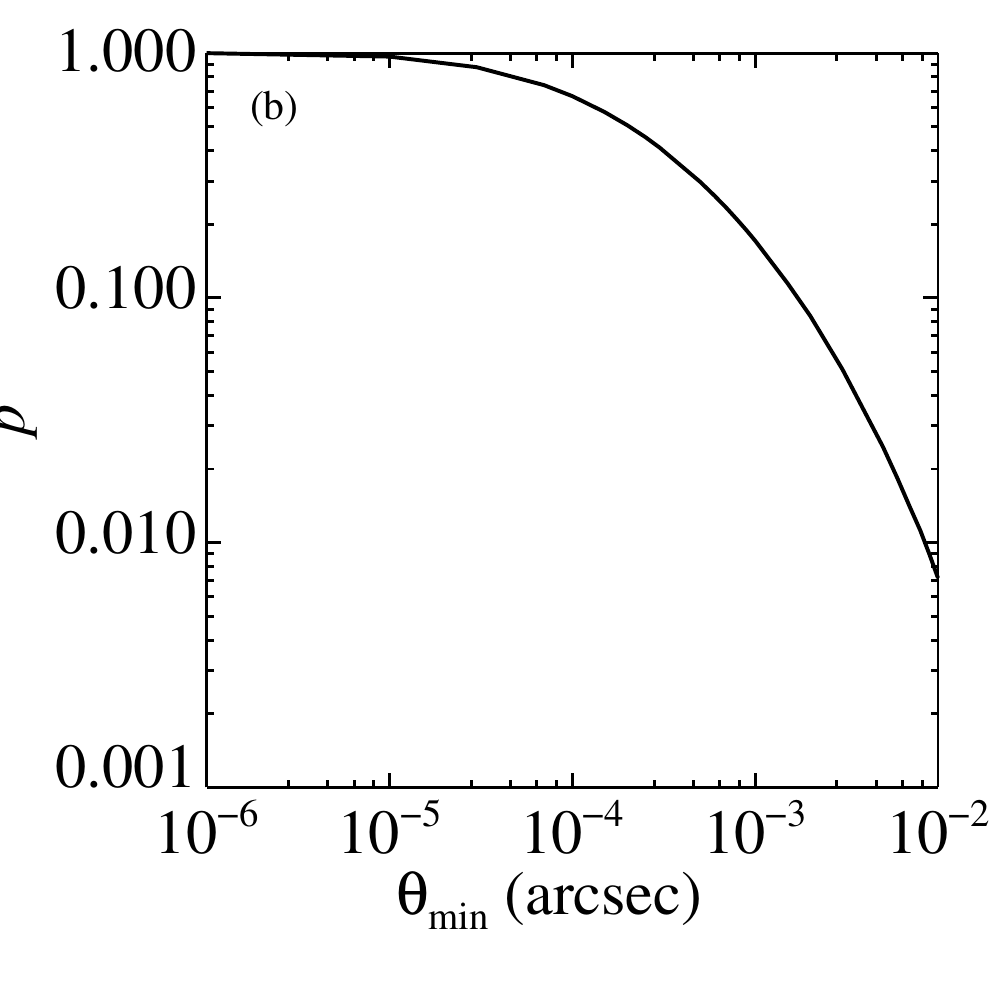}
\includegraphics[width=0.33\textwidth]{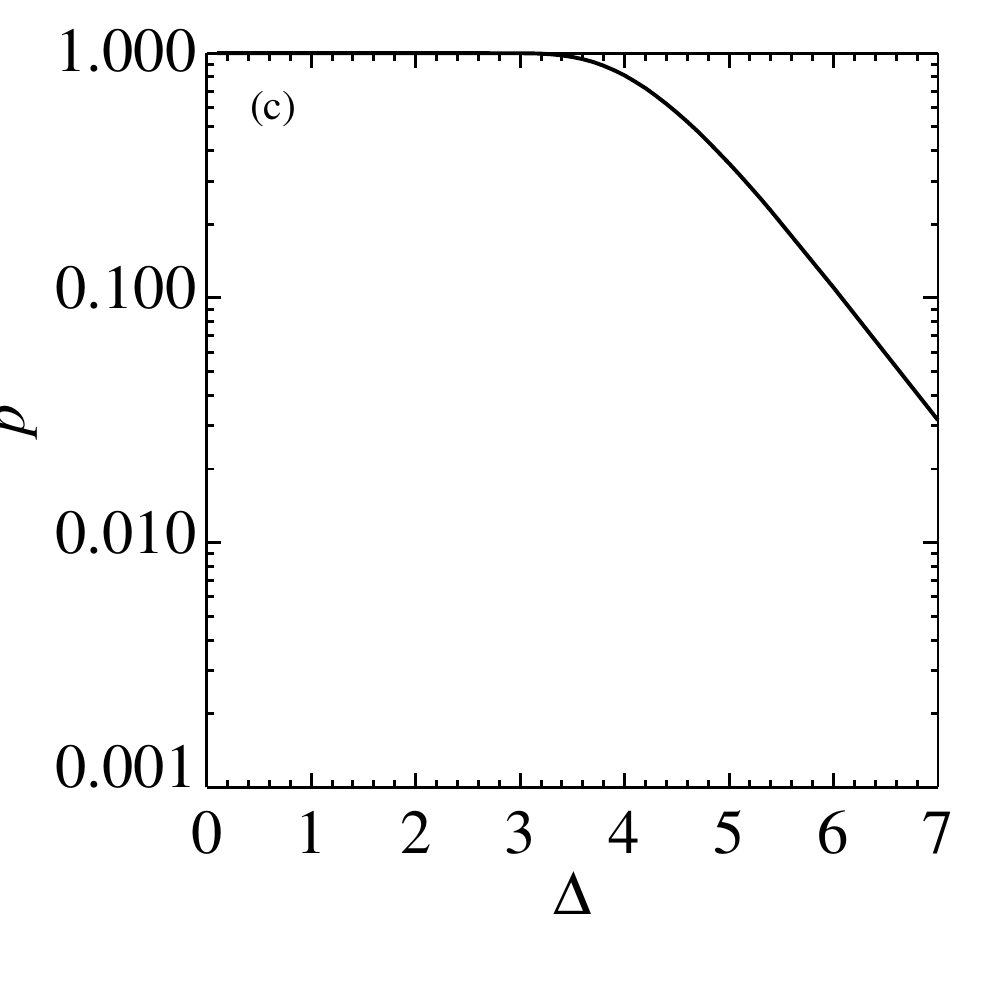}
\caption{Plot of the probability $p$ of detecting 15 or more black
  holes in the top 30 galaxies as a function of the following upper-envelope
  model parameters: (\emph{a}) $\mumin=\log_{10}(M_\mathrm{min}/\msun)$ where
  $M_\mathrm{min}$ is the minimum black-hole mass in the upper-envelope
  model. Values of $\mumin<3$ can be rejected at the 90\% confidence
  level.  (\emph{b}) The minimum detectable angular sphere of
  influence $\theta_\mathrm{min}$.  The $p$-values are very low for
  all plausible values of $\theta_\mathrm{min}$, showing that our test
  is insensitive to the exact value assumed.  (\emph{c}) $\Delta$, the
  range of log mass in the upper-envelope model.}
\label{f:pvsmu}
\label{f:pvssoi}
\label{f:pvsdelta}
\end{figure*}
\bookmarksetup{color=[rgb]{0.54,0,0}}
\bookmark[rellevel=1,keeplevel,dest=pvsmu]{Fig \ref*{f:pvsmu}a: probability vs. minimum mass}
\bookmarksetup{color=black}
\bookmarksetup{color=[rgb]{0.54,0,0}}
\bookmark[rellevel=1,keeplevel,dest=pvssoi]{Fig \ref*{f:pvssoi}b: probability vs. resolution}
\bookmarksetup{color=black}
\bookmarksetup{color=[rgb]{0.54,0,0}}
\bookmark[rellevel=1,keeplevel,dest=pvsdelta]{Fig \ref*{f:pvsdelta}c: probability vs. width}
\bookmarksetup{color=black}

We make the simplifying assumption that a black hole can be detected
if and only if its angular sphere of influence exceeds
$\theta_\mathrm{min}=0$\farcs01, a factor of two smaller than the
smallest angular sphere of influence for a published black-hole mass
\citep[NGC 2778;][]{2003ApJ...583...92G}.  This assumption is
conservative, in that a larger value of $\theta_\mathrm{min}$ would
yield results that are even harder to reconcile with the upper-envelope
model (see Figure \ref{f:pvssoi}b). For a given galaxy, with known
distance $D$ and velocity dispersion $\sigma$, there is then a minimum black-hole
mass that can be detected, $M_\mathrm{limit}(\sigma,D)=\sigma^2D\theta_\mathrm{min}/G$.  In the upper-envelope model
as presented by \cite{batch}, logarithmic black-hole masses
$\mu=\log_{10}\left(M/\msun\right)$ are distributed uniformly between
some upper and lower limits $\mumax$ and $\mumin$, so the
probability of detecting a black hole in a given galaxy is
\beq
P_\mathrm{detect}(\sigma, D) = \frac{\mu_\mathrm{max}(\sigma) - \mu_\mathrm{limit}(\sigma, D)}{\mu_\mathrm{max}(\sigma) - \mu_\mathrm{min}}.
\label{e:probdetect}
\eeq
Following \cite{batch} we take $\mumin=1$ (minimum black-hole mass of
$10\,\msun$) and $\mumax$ given by equation (\ref{eq:msig}) with
$\alpha = 8.7$ and $\beta = 5.0$.  (The assumed values of $\alpha$ and
$\beta$ do not strongly influence our results, and the conclusions
below are made stronger if values that more closely match the best-fit
ridgeline relation are used.)  We use this model to calculate the
probability of making as many black-hole detections as are found in
the top 30 galaxies of Table \ref{t:gallist}.  Of these 30 galaxies,
15 have black-hole mass detections and 15 have either an upper limit
or no published results from a black-hole search.  We make the
conservative assumption that the latter galaxies have no published
detection because their black holes are too small
($\theta_\mathrm{infl}<\theta_\mathrm{min}$).  This is unlikely to be
the case; for example, some of these are unpromising galaxies to
examine for black holes on account of their peculiar or irregular
classifications.

To quantify the probability $p$ of detecting at least 15 black holes,
we ran Monte Carlo realizations of simulated observations of the top
30 galaxies in Table \ref{t:gallist}, using equation
(\ref{e:probdetect}) to calculate detection probabilities of each
galaxy.  Thus, the null hypothesis of this numerical experiment is the
upper envelope model.  For each realization we simulated an
observation of each of the 30 galaxies by drawing a uniform random
number in the range $\left[0,1\right)$.  If the random number was
smaller than $P_\mathrm{detect}(\sigma, D)$, then the black hole was
considered to be detected.  If the null hypothesis, the upper envelope
model, is incorrect, then the numerical experiment should produce very
few realizations in which there were as many or more detections as
there are in the data.  We ran $10^6$ realizations, but only 7166
resulted in 15 or more simulated black-hole detections.  The
probability (or $p$-value) of detecting 15 out of 30 galaxies is then
$p = 7.2\times10^{-3}$, allowing us to reject the null hypothesis
(Batcheldor's version of the upper-envelope model) at the 99.3\%
confidence level.

The above calculation assumes that the minimum logarithmic black-hole
mass is $\mumin = 1$, which was chosen for consistency with
\cite{batch}.  We repeat our simulations for larger values of $\mumin$
and plot $p$ as a function of $\mumin$ in Figure \ref{f:pvsmu}a. Even
for $\mumin=3$ (minimum mass of $1000\,\msun$) we can reject the upper
envelope model at about the 90\% confidence level.

We also repeat our simulations for different values of
$\theta_\mathrm{min}$.  We plot $p$ as a function of the minimum
detectable angular sphere of influence $\theta_\mathrm{min}$ in Figure
\ref{f:pvssoi}b.  Only at $\theta_\mathrm{min} <$0\farcs0015 is $p >
0.1$; thus for all plausible values of $\theta_\mathrm{min}$ the upper
envelope hypothesis can be rejected at high confidence.

Our results are insensitive to the number $N$ of galaxies in our
sample ($N=30$). For $N=20$, $N=30$, and $N=50$ and our standard
parameters ($\mumin=1$ and $\theta_\mathrm{min}=0\farcs01$) the upper-envelope
model is ruled out at the 99.7\%, 99.3\%, and 96.5\% confidence levels
respectively.

We stress again the conservative nature of the assumption that all
galaxies without published black hole mass measurements have black
holes with masses too small to be measured.  If, at the other extreme,
we only considered the 20 galaxies with published mass estimates or
upper limits, then we would rule out Batcheldor's version of the upper-envelope model at
the 99.998\% confidence level or at the 90\% level for $\mumin = 5.1$.

An alternative to the assumption that $\mumin$ is constant is to
assume constant width of the \msigma\ relation so that $\mumin =
\mumax - \Delta$ with a constant value for $\Delta$.  In Figure
\ref{f:pvsdelta}c we plot $p$ as a function of $\Delta$.  The
probability of finding 15 black holes in our sample is $p<0.1$ for
$\Delta>6$.  This result is sensitive to our assumed value for the
normalization parameter $\alpha$ of the \msigma\ relation, taken to be
$\alpha=8.7$ following Batcheldor.  For example, if instead we assume
$\alpha = 8.12$ from G09, we find the much more stringent constraint
$p < 0.1$ for $\Delta > 2.8$.  In order to better constrain this
alternative model as well as to determine the extent that selection
effects alter our inferences, we present a more sophisticated test.

\subsection{Test B}

\label{sec:testb}

We construct a parametrized model for the distribution of black-hole
masses and the observational constraints on their detection. The model
has seven free parameters
$X\equiv\{\alpha,\beta,b,s,\Delta,x_r,s_r\}$, and is based on the
following assumptions:

(\emph{i}) The probability that a given galaxy has a central black
  hole is $b_1$. The parameter $b_1$ is assumed to be independent of
  galaxy properties. This is almost certainly an oversimplification
  but the galaxies in Table \ref{t:gallist} are mostly luminous
  ellipticals, lenticulars, and early-type spirals so are likely to
  have similar properties.

(\emph{ii}) The probability that a galaxy in Table \ref{t:gallist} has been
  examined carefully for evidence of a black hole is some constant $b_2$. Only
  the product $b\equiv b_1b_2$---the combined probability that a
  galaxy has a black hole {\em and} has been examined for one---can be
  determined from the data.

(\emph{iii}) If a galaxy has a black hole, the probability
  distribution of its logarithmic mass
  $\mu=\log_{10}\left(\mbh/\msun\right)$ is determined by the \msigma\
  relation (\ref{eq:msig}) and takes the form
  $dp=p_1(\mu|\sigma,X)d\mu$ where
\begin{equation}
  p_1(\mu|\sigma,X)=g\left[\mu-\alpha-\beta\log_{10}(\sigma/200\kms)\right].
\label{eq:pdef}
\end{equation}
The parameters $\alpha$ and $\beta$ are to be fit from the data, and
the function $g$ is assumed to have the form
\begin{equation}
g(x)=k\left\{\begin{array}{cc}\exp(-\half x^2/s^2), & x > 0 \\
                                               1, & -\Delta \le x \le 0 \\
                                               \exp(-\half(x+\Delta)^2/s^2),
                                               & x < -\Delta.
\end{array}\right.
\label{eq:gdef}
\end{equation}
The case $\Delta=0$ corresponds to the usual assumption of a ridgeline
\msigma\ relation with a Gaussian distribution of the residuals in logarithmic
mass. In the case of a large value of $\Delta$, the \msigma\
relation only defines an upper envelope to the range of black-hole
masses. Since $g$ is a probability density, the constant $k$ must be chosen so
that $\int g(x)dx=1$, that is, $k^{-1}=\Delta+\sqrt{2\pi}s$. With this
parametrization the variance in log mass is
\begin{equation}
\epsilon^2=\frac{\Delta^3/12+\sqrt{\pi/8}\Delta^2s+2\Delta
s^2+\sqrt{2\pi}s^3}{\Delta+\sqrt{2\pi}s}.
\end{equation}

(\emph{iv}) The probability that a black hole will be detected depends only
  on the ratio of the angular radius of the sphere of influence to the
  resolution limit of the telescope. Thus the detection probability is
\begin{equation}
p_2(\mu|\sigma,D,\theta_\mathrm{res},X)=f\left[\log_{10}(\theta_\mathrm{infl}/\theta_\mathrm{res})\right].
\end{equation}
All of the detections in Table \ref{t:gallist} are based on
observations with \emph{HST} or telescopes with inferior resolution
(and thus these could easily have been detected at \emph{HST}
resolution); moreover all of the upper limits are from
\emph{HST}. Thus we can assume \emph{HST} resolution for all of the
measurements in this list, $\theta_\mathrm{res}=0\farcs1$, and
henceforth we suppress this argument.  The value of
$\theta_\mathrm{infl}$ is derived from the actual masss of the black
hole, not the predicted mass and is thus independent of telescope
resolution.  We parametrize the detection probability as
\begin{equation}
f(x)=\left\{\begin{array}{cc}   1, & x > x_r \\
                                               \exp[-\half(x-x_r)^2/s_r^2],
                                               & x \le x_r.
\end{array}\right.
\label{eq:log}
\end{equation}
This equation involves two free parameters: $x_r$ is the value of
$\log_{10}\left(\theta_\mathrm{infl}/\theta_\mathrm{res}\right)$ at
which detection of the black hole becomes certain, and $s_r$ is a
measure of the range of logarithmic black-hole mass over which
detection is possible but not certain. We restrict the ranges of these
parameters to $-1<x_r<1$ and $0<s_r<1$. For example, the restriction
$x_r = 1$ reflects the conservative assumption that detection of a
black hole should be certain if the sphere of influence is more
than ten times the resolution of \emph{HST}. Our results are
insensitive to the values chosen for the range of $x_r$ and $s_r$. The particular functional
form in equation (\ref{eq:log}) is chosen so that the integral in
equation (\ref{eq:pdist}) below is analytic, which greatly speeds up
the time-consuming Markov chain Monte Carlo calculations.

\begin{figure}
\hypertarget{mcmc}{}%
\centering
\includegraphics[width=0.99\columnwidth]{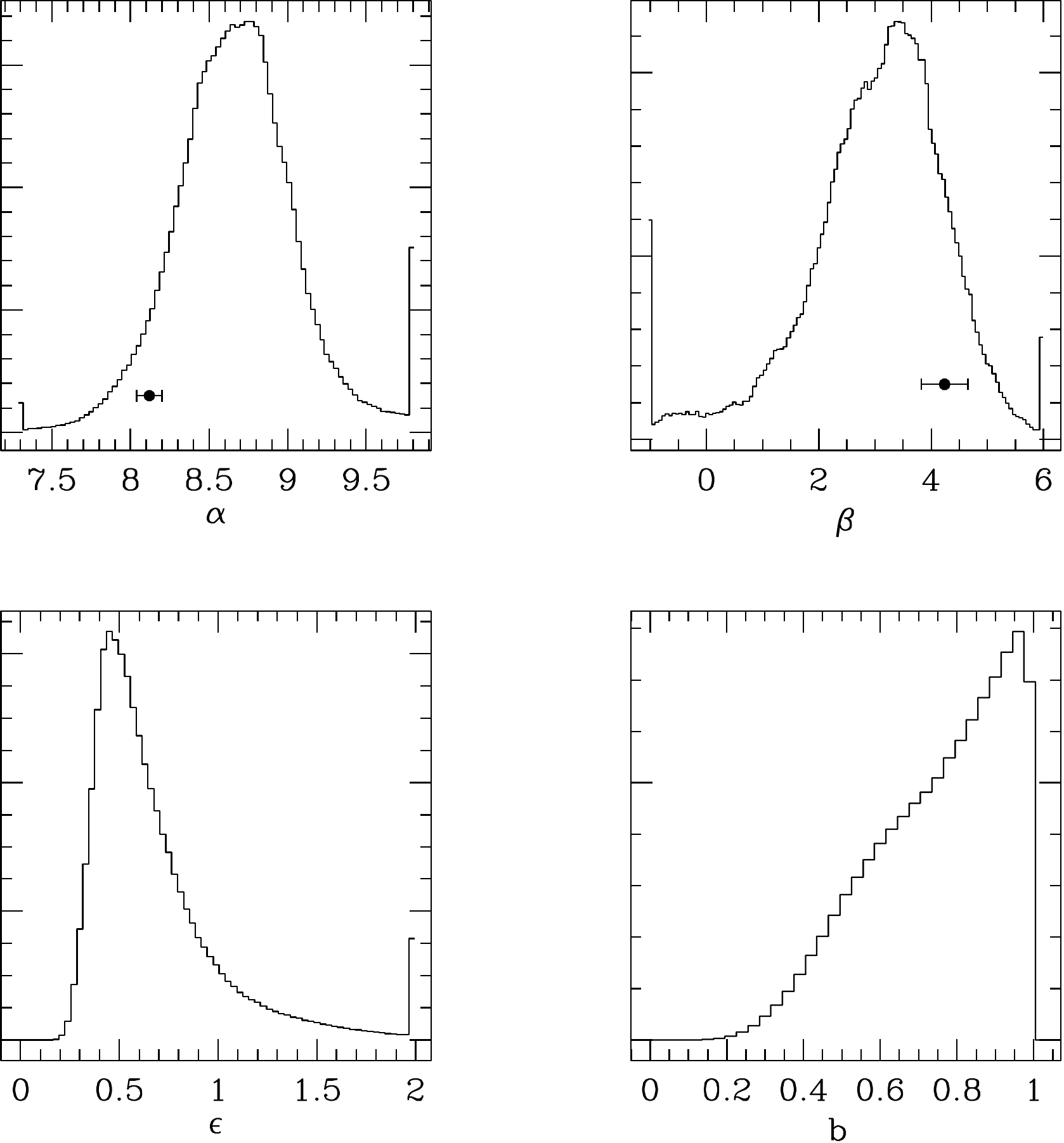}
\caption{The marginalized probabilities of the parameters of the
  \msigma\ relation, as determined by a Markov chain Monte Carlo
  solution of equation (\ref{eq:pdist}) for the data in Table
  \ref{t:gallist}.  The points with error bars denote the estimates of
  normalization $\alpha$ and slope $\beta$ from G09.}
\label{f:mcmc}
\end{figure}
\bookmarksetup{color=[rgb]{0.54,0,0}}
\bookmark[rellevel=1,keeplevel,dest=mcmc]{Fig \ref*{f:mcmc}: MCMC results}
\bookmarksetup{color=black}

Now suppose that we have a sample of $N$ galaxies with dispersions
$\sigma_i$ and distances $D_i$. In $K$ of these galaxies a black hole
has been detected with logarithmic mass $\mu_i$, and in the remaining
$N-K$ galaxies no black hole has been detected (we ignore the extra
information available from the upper limits to the black-hole mass in
a galaxy). Then the posterior probability distribution of the
parameter set $X$ is
\begin{align}
p(X)=&c\,
\Pi(X)b^K\prod_{i=1}^Kp_1(\mu_i|\sigma_i,X)p_2(\mu_i|\sigma_i,D_i,X) \notag \\
&\times \prod_{j=1}^{N-K}
\bigg[1-b\int d\mu p_1(\mu|\sigma_j,X)p_2(\mu|\sigma_j,D_j,X)\bigg],
\label{eq:pdist}
\end{align}
where $\Pi(X)$ is the prior probability distribution and the constant
$c$ is chosen so that $\int p(X)dX=1$.  We assume that the prior
distribution is flat in all of the parameters $X$, with the range
restrictions on $x_r$ and $s_r$ mentioned above. We then evaluate the
probability distribution (\ref{eq:pdist}) using a Markov chain Monte
Carlo simulation and the top $N=30$ galaxies in Table
\ref{t:gallist}. To account for fluctuations in the best-fit
parameters because of the limited sample size, we resample the top 30
galaxies (with replacement) 100 times, run the Markov chain Monte
Carlo each time, and average the results. The marginalized
probability distributions over the parameters $\alpha$ and $\beta$
(the normalization and slope of the \msigma\ relation), $b$ (the
combined probability that a galaxy has a black hole and has been
observed), and $\epsilon$ (the standard deviation in logarithmic mass
of the \msigma\ relation) are shown in Figure \ref{f:mcmc}.

The best-fit values for the parameters of the \msigma\ relation
(\ref{eq:msig}) are $\alpha=8.7\pm0.4$,
$\beta=3.1\genfrac{}{}{0pt}{2}{+1.4}{-1.5}$ (median and 68\% or
1-$\sigma$ confidence interval). These are consistent with the
estimates $\alpha=8.12\pm0.08$, $\beta=4.24\pm0.41$ derived by G09 but
the error bars are much larger and the medians are only consistent at
about the 10\% level. Part of this difference arises because of the
different samples.  The present sample contains only 15 black-hole
masses---less than a third of the 49 masses used  and has a larger
median dispersion ($268\kms$ compared to $175\kms$), both necessary
byproducts of choosing a sample based on
$\theta^\mathrm{pred}_\mathrm{infl}$ (see figure 9b in G09).  Fitting
the masses and upper limits in the present sample using the methods of
G09 yields $\alpha=8.29\pm0.13$, $\beta=3.61\pm0.62$---thus about 30\%
of the difference in the normalization $\alpha$ and about half of the
difference in the slope $\beta$ can be attributed to changes in the
sample. A second reason for the differences is that we are fitting a
more general model---the fit in G09 assumes $\Delta=0$ and ignores
observational selection effects. Our results suggest that accounting
for observational selection may lower the slope and increase the
normalization of the \msigma\ relation from its standard value, but
this is not a secure conclusion since the changes are less than the
1-$\sigma$ confidence interval.

The standard deviation in log mass from the \msigma\ distribution is
$\epsilon=0.6\genfrac{}{}{0pt}{2}{+0.4}{-0.2}$. This is consistent
with the estimate by G09 that $\epsilon=0.44\pm0.06$
for their entire sample and $0.31\pm0.06$ for the ellipticals in their
sample (more precisely, these are estimates of the intrinsic scatter
after removing measurement error, a correction that we do not apply in
this paper).  Thus there is no evidence that the width of the \msigma\
relation derived in prior analyses has been artificially narrowed by
observational selection effects.

\citet{batch} considers models in which the distribution of
logarithmic black-hole mass is uniform over a range of about
6--9. These are similar to models in which the parameter $\Delta$ in
equation (\ref{eq:gdef}) is between 6 and 9. Models in which
$\Delta>6$ are excluded at about the 99\% confidence level.

The probability that a galaxy has a black hole and that it has been
examined for one is $b=0.8\genfrac{}{}{0pt}{2}{+0.15}{-0.2}$. We note
that 20 of the top 30 galaxies in Table \ref{t:gallist} have measured
masses or upper limits, suggesting that the probability that a galaxy
in this list has been examined is $b_2=20/30\approx0.67$. Thus
our results are consistent with $b_1=b/b_2\simeq 1$, i.e., all
galaxies in the list contain central black holes.

A possible concern with this analysis is that our model contains too
many variables to be constrained by the data. One symptom of this
problem would be strong covariances between the model
parameters. We find a significant correlation between $\alpha$ and
$\Delta$ (correlation coefficient 0.5--0.6), which presumably arises
because the mid-point of the ridgeline of the \msigma\ relation at a
given dispersion is determined by the combination $\alpha-\half\Delta$
(eqs.\ \ref{eq:pdef} and \ref{eq:gdef}). There is a strong
anticorrelation between $\alpha$ and $s$ (correlation coefficient
$-0.6$) which presumably arises because the upper envelope of the
\msigma\ relation at a given dispersion is determined by $\alpha+xs$
where $x$ is of order unity. All other correlation coefficients are
typically $\lesssim0.3$ in absolute value. Thus there is no strong
covariance between most of the fitted variables.

The model parameters remain stable as we vary the number of galaxies
in the sample between $N=20$ and $N=50$, The normalization parameter
$\alpha$ declines by only 3\% over this range; the width $s$
increases by about 15--20\%; and the standard deviation in log mass
$\epsilon$ increases by 25--30\%.

Finally, we ask: given the error bars on the parameters of the \msigma\
relation found here, should we believe the smaller error bars from the analysis of
G09? There are good reasons why the error bars in
G09 should be smaller: (\emph{i}) the G09 sample
contains more than three times as many black-hole masses, including some (NGC
4258 and the Milky Way) with very small error bars (this argument assumes, as
did G09, that the \msigma\ relation is the same for these Sbc
spirals as it is for early-type galaxies); (\emph{ii}) the G09 analysis accounts for measurement errors in the mass determinations and
for upper limits; (\emph{iii}) the G09 analysis fits only
three parameters (slope, normalization, and scatter of the \msigma\ relation),
while the present analysis fits seven. Given that the present analysis finds
no evidence for bias due to selection effects in the three parameters that
G09 \emph{do} measure, it is plausible---though not
proven---that such bias is small enough to be negligible.

\section{Discussion and Conclusions}
\label{sec:disc}

The \msigma\ relation was predicted by simple theoretical models of
self-regulated black-hole growth, in which the wind from an accreting
black hole ejects the gas from a galaxy and thereby quenches further
accretion. For an energy-driven wind the predicted relation is
\citep{sr98}
\beq
\mbh = \frac{1}{2 \pi}\frac{\sigma_T}{G^2 m_p c} \frac{f_\mathrm{gas}}{f_w}\sigma^5,
\eeq
where $\sigma_T$ is the Thomson cross section, $m_p$ is the proton
mass, $G$ is the gravitational constant, $c$ is the speed of light,
$f_\mathrm{gas}$ is the gas fraction of the galaxy's total mass, and
$f_w$ is the mechanical power of a wind coming from accretion onto the
black hole, expressed as a fraction of the Eddington
luminosity\footnote{There appears to be an error of a factor of
$(4\pi)^2$ in this formula as given in \citet{sr98}; of course, this is
only an approximate result in any case.}.  For a momentum-driven wind
\citep{1999MNRAS.308L..39F},
\beq 
\mbh = \frac{1}{2\pi}\frac{\sigma_T}{G^2 m_p}\frac{v_w}{c}\frac{f_\mathrm{gas}}{f_{w}}\sigma^4, 
\eeq
where $v_w$ is the wind velocity.  Again, above this mass, all gas is
expelled so that growth by accretion cannot continue unless an
additional source of gas is provided, e.g., by a
merger. Momentum-driven winds are favored because energy-driven winds
appear to be too weak once cooling is accounted for \citep{sn10}.

These theories do not, however, predict whether the growth of black
holes should inevitably continue until these limits are reached, or whether
instead the black-hole growth stalls in many galaxies at smaller
masses, i.e., they do not predict whether \msigma\ is a ridgeline or
an upper-envelope relation.

The tests described in this paper provide strong evidence that
\msigma\ is a ridgeline relation.  In particular, Test A shows that
the upper-envelope relation advocated by \cite{batch} is ruled out because
it predicts far fewer black-hole detections than are found in the
literature.  Quantitatively, our standard upper-envelope model (flat
distribution in log mass down to $10\,\msun$, minimum detectable
sphere of influence $0\farcs01$, $N=30$ galaxies) is inconsistent with
the data at the 99\% level, and the upper-envelope model is ruled
out at $>90\%$ confidence for a wide range of other assumptions.  Test
B shows that after accounting for observational selection effects the
rms scatter in log mass at given dispersion $\sigma$ is only
$\epsilon=0.6\genfrac{}{}{0pt}{2}{+0.4}{-0.2}$, consistent with the
estimate of G09 and inconsistent with the upper-envelope model.

The analysis in Test B also provides a framework for estimating the
bias introduced into the \msigma\ relation by observational selection
effects. Our principal findings are that (\emph{i}) the scatter in the
\msigma\ relation remains small after accounting for observational
selection; (\emph{ii}) the normalization and slope of the relation are
consistent with those derived in analyses that neglect selection
effects, such as G09, but with much larger uncertainties.  These
uncertainties could be reduced by (\emph{i}) searching carefully for
black holes in all of the high-ranked galaxies in Table
\ref{t:gallist} using \emph{HST} or ground-based adaptive optics;
(\emph{ii}) generalizing the analysis to include the black holes in
the Milky Way and those in maser galaxies
\citep{2010arXiv1007.2851G}---of course, a danger in the second step
is that the \msigma\ relation may depend on galaxy morphology.

Finally, our analysis suggests that most galaxies in the list in Table
\ref{t:gallist} do contain a central black hole; in particular, a lower
limit to the probability that a black hole is present---assuming {\em
all} of the galaxies in the list have been searched---is
$b=0.8\genfrac{}{}{0pt}{2}{+0.15}{-0.2}$.

The distribution of black-hole masses as a function of host-galaxy
properties is relevant to the demographics of active galactic nuclei
(AGN), since black holes are believed to be the engines that power
AGN.  In particular, the famous \citet{1982MNRAS.200..115S} argument
estimates the local mass density in black holes from the density of
AGN photons determined from quasar surveys at optical and X-ray
wavelengths. The estimate is based on an assumed radiative efficiency
$\epsilon$ and the ratio of bolometric radiative energy emitted by an
AGN to the rest-mass energy of fuel consumed. The So\l tan density can
be compared to the local density of black holes determined from the
density of galaxies as a function of velocity dispersion and the
\msigma\ relation.  For plausible estimates of the radiative
efficiency (typically $\epsilon=0.1$--0.3 for thin-disk accretion onto
a black hole) these two independent estimates for the black-hole mass
density agree within a factor of two or so
\citep{mr04,2008ApJ...689..732Y}. These results assume a ridgeline
\msigma\ relation so the agreement suggests that the ridgeline model
is not far from correct. This is not a strong argument because of
several uncertain factors in the So\l tan argument such as the
radiative efficiency, the bolometric corrections, and the population
of black holes ejected from galaxy centers by gravitational-wave
recoil.  It is also possible to account for the agreement by invoking
frequent super-Eddington accretion from relatively underweight black
holes \citep{2010MNRAS.408L..95K}, although this hypothesis requires
an active fraction of nuclei much higher than is observed.
Nevertheless, it would be a surprising coincidence if a combination of
errors accidentally canceled in such a way that the simple estimates
we have described for the local black-hole mass density agreed so
well.

Our estimate that the fraction of galaxies in our data set containing
massive black holes is consistent with unity ($b_1 \approx 1$) sheds
light on the process of gravitational-wave recoil in black-hole
mergers.  As two black holes inspiral and coalesce, asymmetric
emission of gravitational waves imparts a kick to the merged black
hole \citep[e.g.,][]{1983MNRAS.203.1049F, 2008ApJ...682L..29B,
2010ApJ...719.1427V}. If this kick is larger than the escape velocity
at the galaxy center, as can happen for high black-hole spins and
particular orientations, then the merged black hole will be ejected.  If the
merging galaxies are typical gas-poor ellipticals,
there will not be enough cold gas at their centers to fuel the growth
of another black hole and re{\"e}stablish the \msigma\ relation.  Our
results therefore suggest that ejection of black holes is rare in
galaxies of this kind, a result consistent with theoretical
calculations \citep{2007ApJ...667L.133S, 2008MNRAS.384.1387V}.
 
We note that because we select galaxies based on
$\theta_\mathrm{infl}^\mathrm{pred}$, our sample tends to have high
velocity dispersions---the median dispersion of the top 30 galaxies in
Table \ref{t:gallist} is $\sigma=268\kms$ compared to a median
$\sigma=175\kms$ in the sample of G09.  Thus the conclusions that we
draw may not apply to the black holes in low-dispersion early-type
galaxies, which are still poorly understood
\citep[e.g.,][]{2010MNRAS.404.2143V}.  There are also observational
hints that late-type and/or small galaxies such as NGC 1068, Circinus,
NGC 4435, and the Milky Way lie systematically below the \msigma\
relation seen in large, early-type galaxies \citep[e.g.,
G09;][]{2010ApJ...721...26G, 2010ApJ...723...54K,
2011Natur.469..374K}.

\hypertarget{ackbkmk}{}%
\acknowledgements 
\bookmark[level=0,dest=ackbkmk]{Acknowledgements}

We thank Tod Lauer and Karl Gebhardt for reading an early draft.

KG acknowledges support provided by the National Aeronautics and Space
Administration through Chandra Award Number GO0-11151X issued by the
Chandra X-ray Observatory Center, which is operated by the Smithsonian
Astrophysical Observatory for and on behalf of the National
Aeronautics Space Administration under contract NAS8-03060 and thanks
the Aspen Center for Physics for their hospitality.
ST acknowledges support from NASA grant NNX08AH24G and NSF grant
AST-0807432.

This research has made use of the NASA/IPAC Extragalactic Database
(NED) which is operated by the Jet Propulsion Laboratory, California
Institute of Technology, under contract with the National Aeronautics
and Space Administration.  We acknowledge the use of the HyperLeda
database (http://leda.univ-lyon1.fr).

\bibliographystyle{apjads}
\hypertarget{refbkmk}{}%
\bookmark[level=0,dest=refbkmk]{References}
\bibliography{gultekin_plus}

\label{lastpage}
\end{document}